\documentstyle[sprocl,psfig]{article}

\def\Journal#1#2#3#4{{#1} {\bf #2}, #3 (#4)}

\def\NPB{{\em Nucl. Phys.} B}
\def\PLB{{\em Phys. Lett.}  B}

\def\PRD{{\em Phys. Rev.} D}

\def\be{\begin{equation}}
\def\ee{\end{equation}}
\def\bea{\begin{eqnarray}}
\def\eea{\end{eqnarray}}

\begin{document}

\title{LATTICE QCD WITH THE OVERLAP-DIRAC OPERATOR: ITS $\Lambda$
PARAMETER, AND ONE-LOOP RENORMALIZATION OF FERMIONIC CURRENTS}

\author{C. ALEXANDROU, E. FOLLANA, H. PANAGOPOULOS\footnote{Presented
the talk}}

\address{Department of Physics, University of Cyprus,\\
P.O.Box 20537, Nicosia CY-1678, Cyprus}

\author{E. VICARI}

\address{Dipartimento di Fisica dell'Universit\`a 
and I.N.F.N.,\\
Via Buonarroti 2, I-56127 Pisa, Italy}


\maketitle\abstracts{We compute the ratio between the scale $\Lambda_L$ 
associated with a lattice formulation of QCD using the overlap-Dirac
operator, and $\Lambda_{\overline{MS}}$. To this end, the
one-loop relation between the lattice 
coupling $g_0$ and the coupling renormalized in the $\overline{{\rm MS}}$ scheme is
calculated, using the lattice background field technique. 
We also compute the one-loop
renormalization $Z_\Gamma$ of the two-quark operators $\bar{\psi} \Gamma \psi$, where
$\Gamma$ denotes a generic Dirac matrix. Furthermore, we study the
renormalization of quark bilinears which are more extended and have better
chiral properties.
Finally, we present improved estimates of $Z_\Gamma$,
coming from cactus resummation and from mean field perturbation theory.}

It has recently been shown that chiral symmetry can be
realized in lattice QCD without fermion doubling, circumventing 
the Nielsen-Ninomiya theorem (for a list of
references, see our publications [1,2]; for reviews see,
e.g., Refs. [3,4]).
This has been achieved by introducing an overlap-Dirac operator derived 
from the overlap formulation of chiral fermions~\cite{N-N-95}.
The simplest such example, for a massless fermion, is given by
the Neuberger-Dirac operator~\cite{Neuberger-98}
\be
D_{\rm N} = {1\over a} \,\rho \left[  1 + X (X^\dagger X)^{-1/2}
\right], \qquad X= D_{\rm W} - {1\over a}\rho,
\label{Nop}
\ee
$a$ is the lattice spacing, $\rho{\in}(0,2)$ a parameter,
$D_{\rm W}$ the Wilson-Dirac operator
\be
D_{\rm W} = {1\over 2} \left[ \gamma_\mu \left( \nabla_\mu^*{+}\nabla_\mu\right)
 {-} a\nabla_\mu^*\nabla_\mu \right], \ \ 
a\nabla_\mu\psi(x) = U(x,\mu) \psi(x + a\hat{\mu}){-} \psi(x).
\ee

$D_{\rm N}$ has a number of desirable features: The
Ginsparg-Wilson relation:\\
$\gamma_5\, D_{\rm N} + D_{\rm N}\,\gamma_5
= a \,D_{\rm N}\,\gamma_5 \,D_{\rm N},$
protects the quark masses from 
additive renormalization, and implies renormalizability to all orders 
of perturbation theory~\cite{R-R-99}.
This relation also leads to the existence of an exact chiral symmetry
of the lattice action~\cite{Luscher-98},
with chiral Ward identities which ensure the
non-renormalization of vector and flavor non-singlet axial vector
currents and the absence of mixing among operators in different chiral
representations.
Chiral symmetry results in leading scaling corrections to hadron
masses which are $O(a^2)$, rather than $O(a)$.
The axial anomaly is correctly reproduced by the fermion integral
measure, which is non-invariant under flavour-singlet chiral 
transformations.
$D_{\rm N}$ avoids fermion doubling at the expense of not
being strictly local:
Locality is recovered in a more general sense, i.e. allowing 
an exponential decay of the kernel of $D_{\rm N}$ at a rate which scales
with the lattice spacing and not with the physical quantities~\cite{H-J-L-99}.

In what follows, we present perturbative calculations, in
lattice QCD with the operator $D_{\rm N}$, of several quantities
which are needed to relate Monte Carlo data to physical
observables. Lack of strict locality greatly complicates these
calculations, as compared to the Wilson case. Technical details may be
found in our publications [1,2].

To evaluate $\Lambda_{\rm L}/\Lambda_{\overline{\rm MS}}$
we need to calculate~\cite{apv} the one-loop relation between $g_0$ and
the renormalized  
$\overline{\rm MS}$ coupling $g$ at scale $\mu$: $g_0 = Z_g(g_0,a\mu)
g$. Writing:\\
$Z_g(g_0,x)^2  {=} 1 {+} g_0^2\left( 2 b_0 \ln x {+} l_0 \right)
{+} {\cal O}(g_0^4),$ one has: $l_0 {=} 2 b_0\ln
\left({\Lambda_L/\Lambda_{\overline{\rm MS}}}\right).$

The algebra was performed using a
symbolic manipulation package which we have developed in Mathematica. 
For the present purposes, this package was
augmented to include the propagator and vertices of the overlap action.

Our results are shown in Figure 1 for different numbers $N_f$ of fermion flavours.
Some particular cases of interest are ($SU(3),\ \rho{=}1$):\\
$\Lambda_L/\Lambda_{\overline{\rm MS}} = 0.034711\ (N_f{=}0),\quad 
0.025042\ (N_f{=}1), \quad 0.011273\ (N_f{=}3),$\\
(cf. Wilson fermions:
$\Lambda_L/\Lambda_{\overline{\rm MS}} = 0.029412\ (N_f{=}1),\quad
0.019618\ (N_f{=}3)\,).$

\bigskip
\noindent
\begin{minipage}{5.8truecm}
\psfig{figure=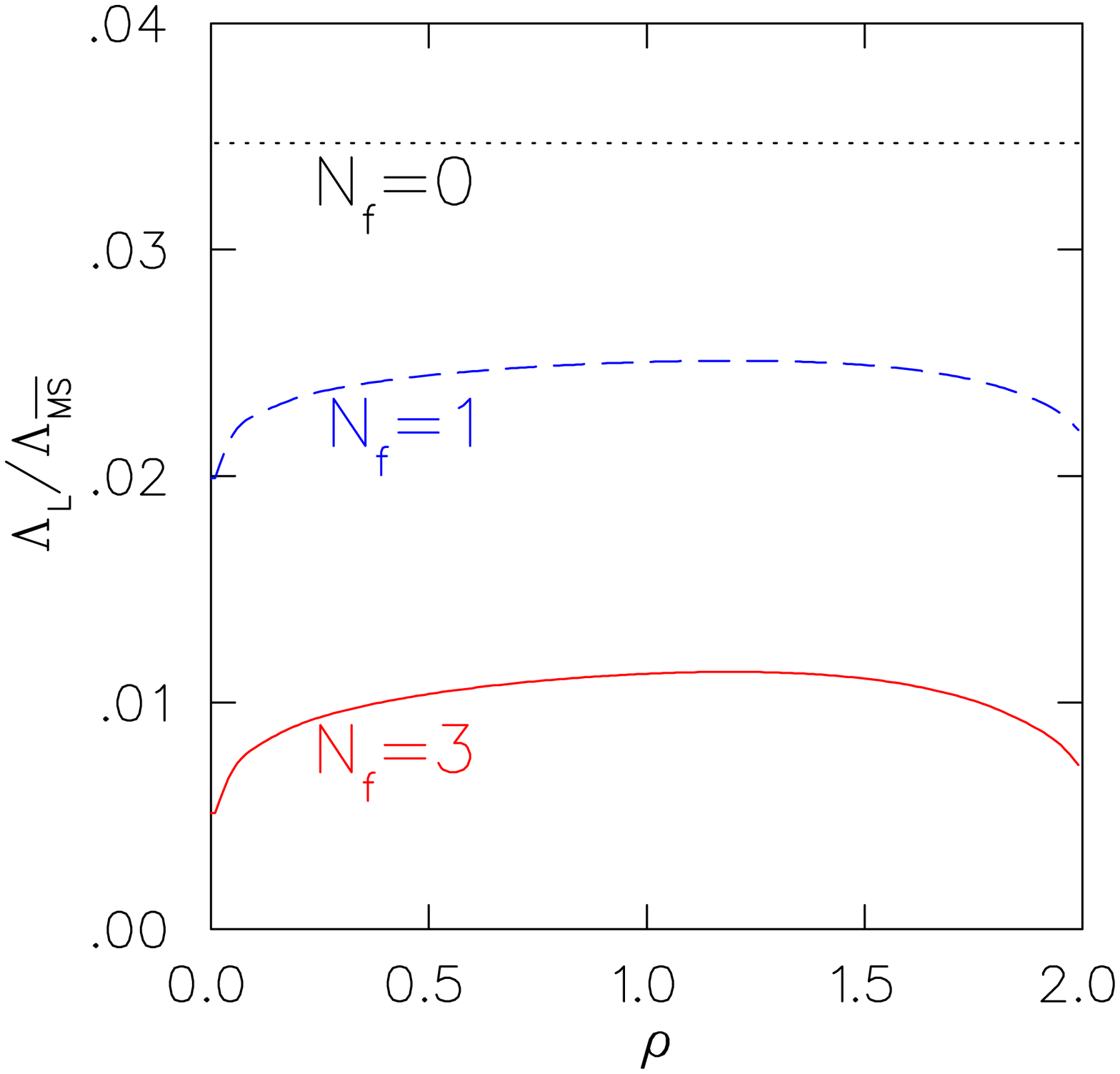,width=5.8truecm}
\footnotesize Figure 1: $\Lambda_L/\Lambda_{\overline{\rm MS}}$ in $SU(3)$,
as a function of $\rho\,$.
\end{minipage}\hskip0.4truecm
\begin{minipage}{5.8truecm}
\psfig{figure=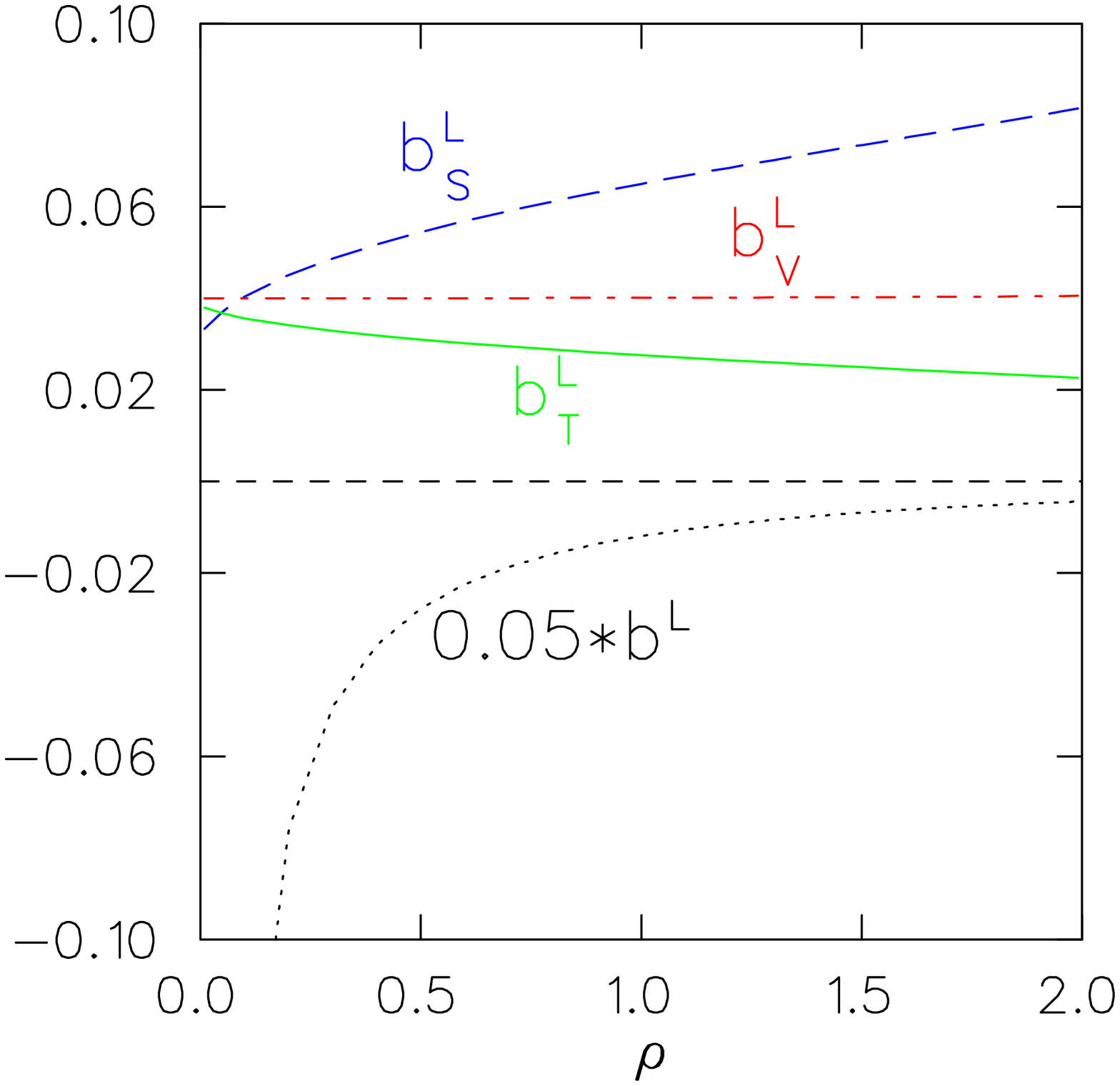,width=5.8truecm}
\footnotesize Figure 2: The coefficients $b^{\rm L}(\rho), b^{\rm L}_{O_i}(\rho)$,
as a function of $\rho\,$.
\end{minipage}

We have furthermore computed~\cite{afpv}, to one loop,
the renormalization constants $Z_O$ of the local fermionic currents:

$O_i = \bar{\psi}(x) \Gamma_i \psi(x), \qquad \Gamma_i = 
1\ (S),\,\gamma_5\ (P),\,\gamma_\mu\ (V),\,\gamma_\mu\gamma_5\
(A),\,\sigma_{\mu\nu}\gamma_5\ (T),$\\
and their extended (non-ultralocal), improved counterparts: 

$O'_i \equiv \bar{\psi} \Gamma_i \left( 1 - a D_{\rm
N}/2\right) \psi,\quad
O''_i \equiv \bar{\psi} \left( 1 - a D_{\rm N}/2\right) \Gamma_i
\left( 1 - a D_{\rm N}/2\right) \psi.$\\
($O'_i$ obey Ward identities leading to: $Z_{S'}=Z_{P'},\ 
Z_{V'}=Z_{A'}.$ $O''_i$ are free of ${\cal O}(a)$ errors,
not only in the spectrum, but also in generic matrix elements.)

We have proved that: $Z_{O'_i} = Z_{O_i}\,$, and also: $Z_{O''_i} = Z_{O_i}\,$.

We calculated $Z_{O_i} = 1 + g^2 c_F
[(c_{O_i} - c) \ln a^2 \mu^2 +
b^{\overline{\rm MS}}_{O_i} - b^{\overline{\rm MS}}-b^{\rm L}_{O_i} +
b^{\rm L}]$ (see Ref. [2] for notation).
$Z_{O_i}$ are independent of the gauge parameter
and of the fermion mass. The results for $b^{\rm L}$ and $b^{\rm
L}_{O_i}$ are shown in Figure 2; they do not depend on $N$ or $N_f$.
As an example, $Z_{O_i}$ at $\rho=1$ is:\\
$Z_{S,P} = 1 + g^2 c_F \bigl[\, 3\, (\ln a^2 \mu^2)/16\pi^2 + 0.204977
\,\bigr],$
$Z_{A,V} = 1 + g^2 c_F \bigl[\, 0.198206  \,\bigr],$\\
$Z_{T}   = 1 + g^2 c_F \bigl[\, - (\ln a^2 \mu^2)/16\pi^2 + 0.204392 \,\bigr].$

\medskip
Finally, we have obtained improved estimates for $Z_{O_i}$, coming from a
resummation to all orders of ``cactus'' diagrams~\cite{cactus}.
These diagrams are often largely responsible for
lattice artifacts. Our method is gauge invariant, and systematic in dressing
higher loop contributions; applied to a
number of cases of interest, it has yielded results remarkably close to 
nonperturbative estimates. 

In particular, for $Z_{V,A}$ we find (at $g_0 = 1,\ \rho=1$): 
$Z_{V,A} \simeq  1.35$, as compared to our undressed result: $Z_{V,A}
= 1.26427\,.$

To conclude, some {\it feasible} future tasks, alongside
with numerical simulation, are:
Calculation of the $\beta$-function for the overlap-Dirac operator,
running fermion masses, renormalization of 4-fermion
operators~\cite{Capitani} ($Z^{\rm \Delta S{=}2}$, etc.).

\end{document}